\documentclass{acm_proc_article-sp}

\usepackage{booktabs} 
\usepackage{url, graphicx, subcaption, enumitem}
\usepackage{makecell}
\usepackage{color}

\setlist{nolistsep}

\usepackage{etoolbox}
\makeatletter
\patchcmd{\maketitle}{\@copyrightspace}{}{}{}
\makeatother

\begin{document}

\conferenceinfo{}{Bloomberg Data for Good Exchange 2018, NY, USA}

\title{Financial Forecasting and Analysis for Low-Wage Workers}

\numberofauthors{7}
\author{
\alignauthor
Wenyu Zhang\thanks{This work was conducted under the auspices of the IBM Science for Social Good initiative.}\\
       \affaddr{Cornell University}\\
       \affaddr{Ithaca, NY}\\
       \email{wz258@cornell.edu}
\alignauthor
Raya Horesh\\
       \affaddr{IBM Research}\\
       \affaddr{Yorktown Heights, NY}\\
       \email{rhoresh@us.ibm.com}
\alignauthor
Karthikeyan Natesan Ramamurthy\\
       \affaddr{IBM Research}\\
       \affaddr{Yorktown Heights, NY}\\
       \email{knatesa@us.ibm.com}   
\and
\alignauthor
Lingfei Wu\\
       \affaddr{IBM Research}\\
       \affaddr{Yorktown Heights, NY}\\
       \email{wuli@us.ibm.com}    
\alignauthor
Jinfeng Yi\thanks{Now at Tencent AI Lab, Bellevue, WA, USA.}\\
       \affaddr{IBM Research}\\
       \affaddr{Yorktown Heights, NY}\\
       \email{jinfengyi.ustc@gmail.com}   
\alignauthor
Kryn Anderson\\
       \affaddr{Neighborhood Trust Financial Partners}\\
       \affaddr{New York, NY}\\
       \email{kanderson@neighborhoodtrust.org}         
\and
\alignauthor
Kush R.\ Varshney\\
       \affaddr{IBM Research}\\
       \affaddr{Yorktown Heights, NY}\\
       \email{krvarshn@us.ibm.com}
}

\maketitle

\begin{abstract}

Despite the plethora of financial services and products on the market nowadays,  there is a lack of such services and products designed especially for the low-wage population. Approximately 30\% of the U.S. working population engage in low-wage work, and many of them lead a paycheck-to-paycheck lifestyle. Financial planning advice needs to explicitly address their financial instability.  In this paper, we propose a system of data mining techniques on small-scale transactions data to improve automatic and personalized financial planning advice to low-wage workers. We propose robust methods for accurate prediction of bank account balances and automatic extraction of recurring transactions and unexpected large expenses. We formulate a hybrid method consisting of historical data averaging and a regularized regression framework for prediction. To uncover recurring transactions, we use a heuristic approach that capitalizes on transaction descriptions. Our methods achieve higher performance compared to conventional approaches and state-of-the-art predictive methods in real financial transactions data. The proposed methods will upgrade the functionalities in WageGoal, Neighborhood Trust Financial Partners' web-based application that provides budgeting and cash flow management services to a user base comprising mostly low-income individuals. The proposed methods will therefore have a direct impact on the individuals who are or will be connected to the product.

\end{abstract}

\section{Background and Motivation}

Low-wage workers make up a large portion of the working population, and need the most help in financial planning. The U.S. Bureau of Labor Statistics defines low-wage work according to three hourly-wage levels \cite{fusaro16}:
\begin{itemize}
\setlength\itemsep{0em}
\item (\$9.25) to lift family of two above poverty line,
\item (\$10.75) to lift family of three above poverty line,
\item (\$13.50) to lift family of three to 125\% of poverty line,
\end{itemize}
where the corresponding wages in 2013 are in parenthesis. Approximately 15,000,000 U.S. workers earned up to \$9.25/hr and 36,000,000 earned up to \$13.50/hr in 2013. These are equivalent to 12.63\% and 29.54\% of the U.S. working population respectively. To put the wages into perspective, \$13.50/hr amounts to a monthly income of just over \$2000, while the median rent for one-bedroom apartments across 50 major U.S. cities is already \$1200 \cite{bi16}.

Many low-income households live paycheck-to-paycheck, at risk of financial instability in the event of medical, job or other unforseen emergencies when they are forced to take up loans and in the process incur additional charges. Reference \cite{pew16} reports that the largest U.S. banks charged \$11.6 billion in overdraft and insufficient fund fees in 2015, of which a significant portion is attributed to the poor. To make matters worse, the average debit charge triggering such fees is \$24, while the typical overdraft fee is \$35.

The poor are affected by obstacles including high cost of personalized financial services, lack of suitable banking services \cite{barr04}, and hassle of seeking financial services \cite{mullainathan10}. We hope to alleviate these obstacles through data-driven solutions. In this paper, we propose a system of data mining techniques to improve automatic and personalized financial planning advice to low-wage workers. We work with anonymized checking, savings, and credit card account transactions data. User identification information such as age, gender, and size of household is not used. We also work in the small data scenario where low-income individuals may not have a long banking history due to the aforementioned obstacles.

\begin{figure}
\setlength{\belowcaptionskip}{-10pt}
\centering
\includegraphics[width=\linewidth]{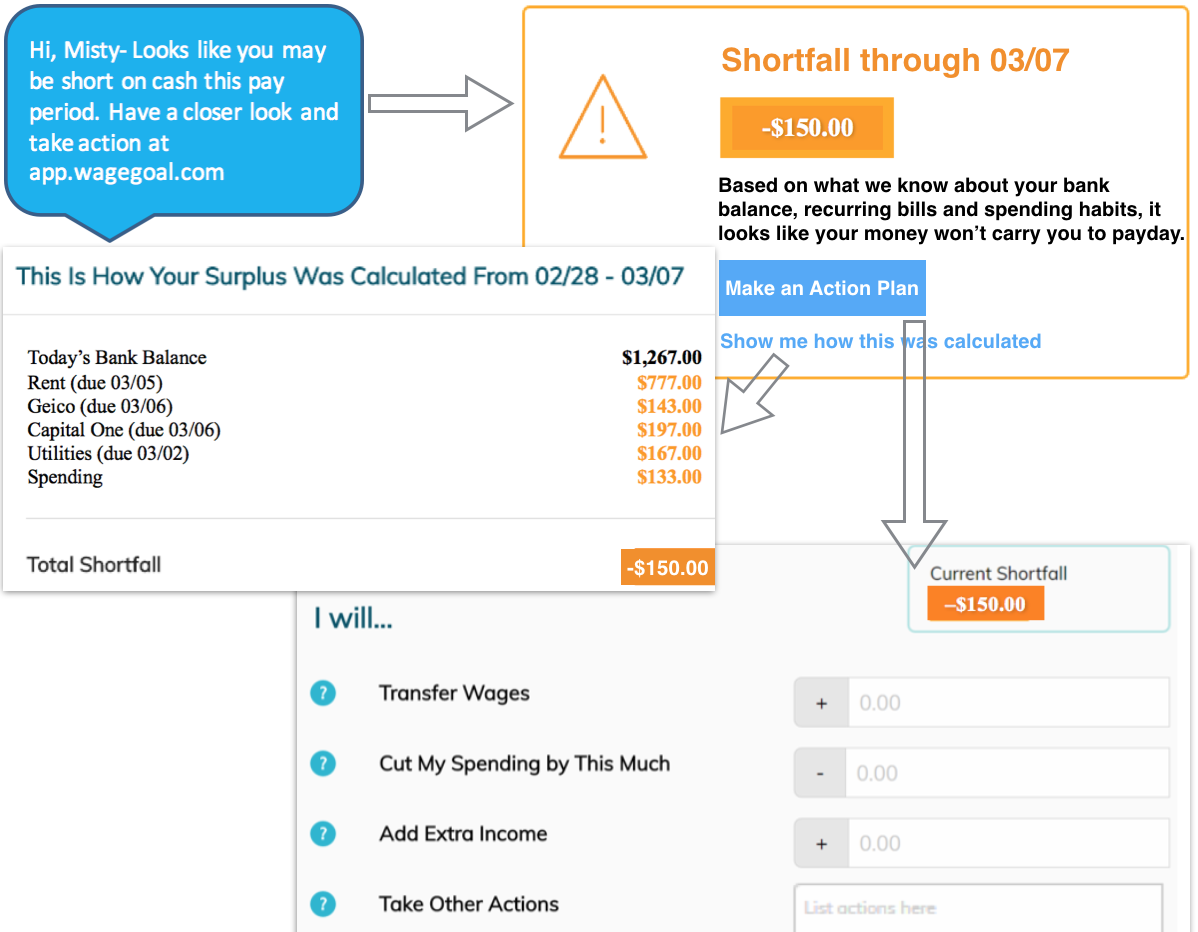}
\caption{Some current functionalities of WageGoal.}
\label{fig: wagegoal}
\end{figure}

\begin{figure}
\setlength{\belowcaptionskip}{-10pt}
\centering
\includegraphics[width=0.9\linewidth]{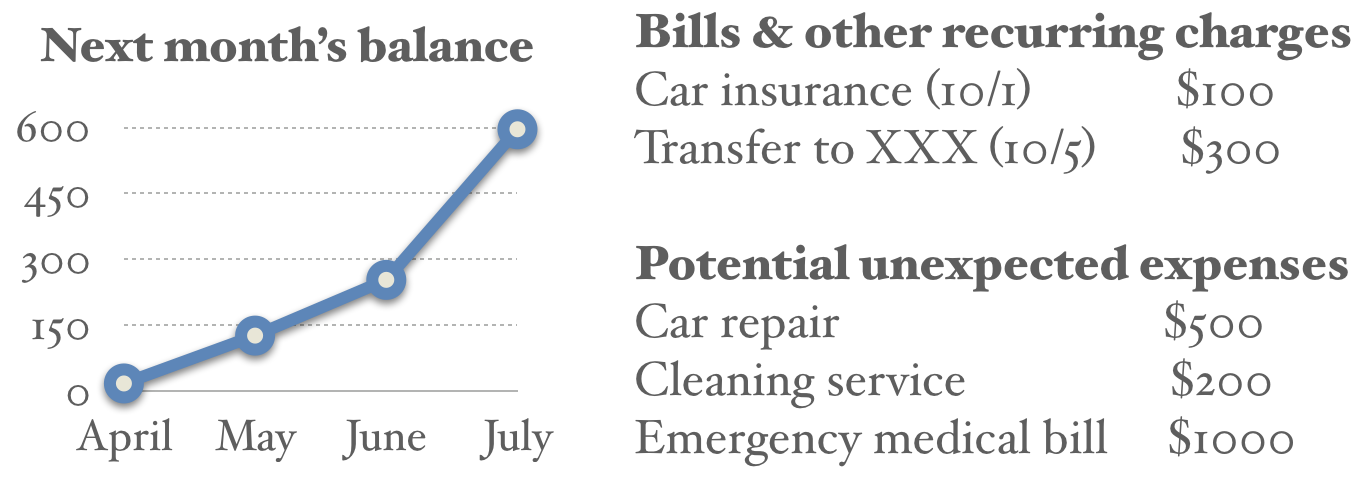}
\caption{Proposed functionalities for WageGoal.}
\label{fig: panel}
\end{figure}

This work is motivated by use cases for Neighborhood Trust Financial Partners' WageGoal app which caters primarily to low-income individuals \cite{nt16}. 
Figure \ref{fig: wagegoal} contains screenshots illustrating some current functionalities such as giving a cash flow snapshot of the user's income, bills and expenses, and helping with cash flow management. As illustrated in Figure \ref{fig: panel}, our main contributions are to propose methods for:
\begin{itemize}
\setlength\itemsep{0em}
\item Short- and long-term prediction of bank account balances with improved accuracy;
\item Automatic extraction of recurring transactions and unexpected large expenses.
\end{itemize}

The functionalities inform users about their possible future spending behavior so that they have enough time to adjust and to save up for emergencies. The main difficulties of these tasks originate from the multifaceted nature of transactions data. A user's spending depends on individual needs and historical spending, but can also exhibit patterns similar to other users. Moreover, salary, bills and other recurring transactions provide characteristic features of a user's spending behavior, but these cyclic patterns can be noisy and inconsistent at times.
Our methods address these difficulties by both effectively mining a user's recurring transactions and borrowing strength from the spending patterns of other users. We tested on two real financial transactions datasets, one is a smaller dataset from actual WageGoal users, and the other is a larger publicly available dataset from PKDD'99 Discovery Challenge. Our methods achieve higher performance compared to conventional approaches and state-of-the-art prediction methods on both datasets.

We differentiate the proposed functionalities from those already offered by personal finance apps on the market \cite{bi17}. The apps mainly track cash flow and provide simple budgeting tools, such as calculating an average daily ``spendable" amount based on a user's income and saving goal. For low-wage workers whose bank balance can hover dangerously around zero, more accurate estimates and finer-grained financial analysis are necessary.

\section{Transactions Data: Overview}
\label{sec: overview}

The WageGoal app collects users' transactions through authorized accounts using a third-party service Plaid \cite{plaid}. Each transaction is associated with an account ID, date, description or merchant name, amount and category. 
Table \ref{table: overview} shows a sample snippet of the transactions captured.
There are a total of 11 categories, namely Bank Fees, Cash Advance, Community, Food and Drink, Healthcare, Interest, Payment, Recreation, Service, Shops, and Travel.
Uncategorized transactions are labeled NA. In the WageGoal Dataset, 28\% of the transactions are uncategorized. The category labels are provided through Plaid and we do not attempt to address the problem of missing labels in this paper. Daily account balances are retrospectively calculated from current balances.

\begin{table}

\centering
\begin{tabular}{*{5}{l}}
\makecell{ID} & Date & Description & \$ & Category \\
\hline
1 & 6/22/2016 & IKEA & 20 & Shops \\
2 & 6/22/2016 & Target & 10 & NA \\
1 & 6/24/2016 & Starbucks & 15 & Food \& Drink \\
3 & 6/24/2016 & Interest & -0.01 & Interest \\
3 & 6/24/2016 & Direct Deposit & -1000 & Transfer
\end{tabular}

\caption{WageGoal Dataset: Examples of transactions data.}
\label{table: overview}
\vspace{-15pt}
\end{table}

\subsection{Initial Cluster Analysis}

The WageGoal dataset consists of 19 users with approximately one year of transactions. Using Dynamic Time Warping (DTW) distance \cite{berndt1994using,wu2018random}, we cluster the overall balance sequences for the last available month by hierarchical agglomerative clustering. Balance sequences are set to zero-mean which does not affect the spending pattern. We use DTW window = 2 to allow for slight time series misalignments.

\begin{figure}
\setlength{\belowcaptionskip}{-2pt}
  \centering
  \begin{subfigure}[b]{\linewidth}
  \centering
  \includegraphics[width=0.35\textwidth]{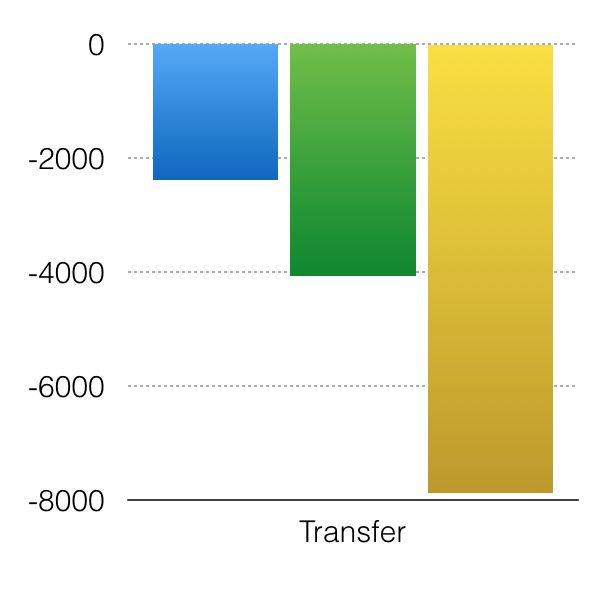}
  \caption{Transfer (negative spending = income)}
  \end{subfigure}
  
  \begin{subfigure}[b]{\linewidth}
  \includegraphics[width=\textwidth]{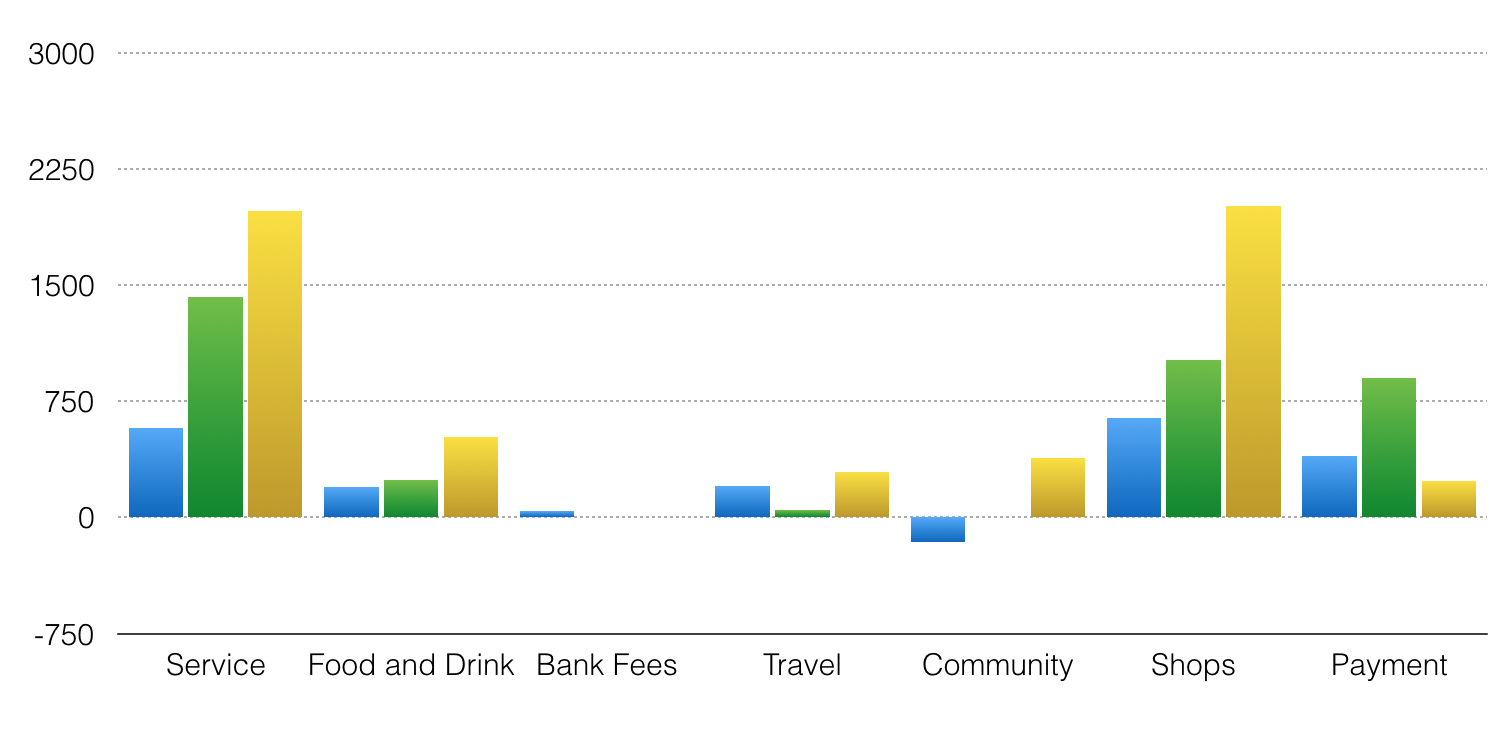}
  \caption{Selected categories}
  \end{subfigure} 
  \caption{WageGoal Dataset: Average category-wise spending for each cluster of users: blue, green and yellow.}
  \label{fig: cat}
  \vspace{-10pt}
\end{figure}

Setting the number of clusters to five, two clusters consist of one user each. For the three other clusters, we plot their average category-wise spending in Figure \ref{fig: cat}, which shows a clear distinction between the less and more well-off users. The least well-off (blue) group spent less and also incurred more bank fees compared to the moderate (green) and the most well-off (yellow) groups.
The three groups spent similarly in four other categories omitted from the figure.

Temporal patterns in balance sequences help to distinguish the users' financial status. Hence, we devise our prediction method to borrow strength from the data of similar users.

\section{Related Work}

Most works on transactions data make use of RFM (Recency, Frequency and Monetary value) to define and analyze customer value \cite{birant11,khajvand11,chang11}. However, these summary statistics mask too many details for our purpose. Instead, we devise data mining and time series techniques to extract more information from the raw data and also make use of similarities amongst users' spending behaviors to effectively improve prediction.

\subsection{Prediction}

Two-part models are popular in modeling household finances, medical expenditure and other nonnegative data \cite{brown15,min02,mullahy98,neelon11}, and can be formulated to cluster the subjects such that each cluster receives different parameter values. They often use covariates as part of the binary and continuous component representations. Covariates include time-related and class-membership variables, such as gender and employee vs. dependency status \cite{neelon11}, which need extra manual effort to construct or are simply not available in anonymized data.

A traditional time series model is the ARMA (autoregressive moving average) which regresses the current variable value on the past \cite{shumway06}. Seasonal ARMA is used on periodic data such as traffic flow \cite{williams03}. Since transactions contain multiple seasonal components which may not have fixed periodicities, a plain seasonal ARMA is not suitable. Neural networks are used to model ARMA model residuals \cite{zhang03} or directly to predict \cite{ruiz16}, but they require large data for good performance.

In other data-driven time series analysis literature, Taylor and Letham \cite{taylor17} implement a fast regression-based method that includes holiday effects, but models only weekly seasonality and requires handcrafted variables to indicate holiday effects. Approaches reported in \cite{alvarez11,scott14,zhang13} extract similar sequences in historical data and use KNN regression to predict by taking unweighted or weighted averages of the samples immediately after the matched sequences. These may not be robust enough against ``anomalies" or spikes not uncommon in expenditure. 
To induce sparsity on regression coefficients, \cite{scott14} uses a spike-and-slab prior. Markov chain Monte Carlo methods estimate the full Bayesian model parameters, but they take significant computation time to iteratively forecast multiple future days, one day at a time. 

While the aforementioned methods are suitable for ``well-behaved" time series, we incorporate design features suitable for transactions, while working under the framework that no further data annotations are required and data is limited at early stages of user enrollment. We recognize that different bank account types and prediction horizons require different treatment, and hence propose a hybrid approach. One aspect of our approach works on the level of individual transactions and relies on extracted recurring transactions, while the other works from the holistic point-of-view of overall balance, where we extract similar sequences and use a computationally efficient regularized regression scheme to penalize anomalous sequences. Another key innovation is that we align the extracted sequences using landmark transactions before regressing, which increases prediction accuracy.

\subsection{Finding recurring transactions}
\label{sec: related recur}

It is common to identify time series periodicities in the frequency domain. 
A Lomb-Scargle periodogram approach to treat missing values and unevenly-spaced time points in finding periodicity in gene expression patterns is proposed in \cite{glynn06}. Another thread line of research \cite{wu2018random,lei2017similarity} learns the vector-form representation for multi-variable time-series to use in subsequent downstreaming tasks. There are also methods such as \cite{elfeky05} which address the problem directly in the time domain. Although some of these methods are capable of detecting multiple periodicities, they require the period of each cyclic component to be consistent across time. In transactions, there can be significant jitter in the periodicity for recurring transactions such as payments due to differences in the number of days each month and the presence of holidays. Moreover, just using numerical values is insufficient to identify recurring transactions since there is substantial noise, and this is especially so for recurring transactions with small dollar amounts. Manually constructing and maintaining a complete biller's list for each user's recurring transactions is ideal but tedious. Hence in this paper, we propose a procedure that automatically identifies possible recurring transactions that takes into account the inconsistent nature of transactions and better distinguish between transactions through their textual descriptions.

\section{Technical Solutions}

We propose methods that use transactions data as described in Section \ref{sec: overview}. The main challenges of working with transactions data versus conventional numerical time series are the presence of:
\begin{itemize}
\setlength\itemsep{0em}
\item Text description for each transaction; 
\item  Multiple noisy and inconsistent periodic patterns;
\item Spikes in spending.
\end{itemize}

\subsection{Prediction of account balances}
\label{sec: pred}

We predict account balances up to 31 days ahead to encompass two semimonthly paydays to give users sufficient time and information to plan their finances.
We propose a historical averaging method \emph{HistAvg} for short-term prediction and accounts with minimal transactions, and a regularized least squares method \emph{SubseqLS} for long-term prediction of accounts with distinct cyclic patterns. To effectively address the nuances in modeling different account types, we further propose a hybrid method \emph{HistAvg-SubseqLS} where the first $\tau$ days are predicted by \emph{HistAvg} and the rest by \emph{SubseqLS}.

\subsubsection{HistAvg}

HistAvg predicts daily spending and is adapted from the current implementation in WageGoal. The original version uses a biller's list for bills, while we use recurring transactions found by our proposed procedure in Section \ref{sec: recur}. Predicting spending using past three months' transactions follows:
\begin{enumerate}
\setlength\itemsep{0em}
\item Remove recurring charges and top 10\% of transactions;
\item Calculate daily basic spending as the average amount spent daily according to the remaining transactions;
\item Estimate future spending as the sum of daily basic spending and any recurring charge on that day.
\end{enumerate}

Top 10\% of transactions are excluded since these are typically rare purchases. Finally, the account balance is the sum of the previous day's balance and the estimated spending.

\subsubsection{SubseqLS}
\label{sec: SubseqLS}

We assume that a similar balance history implies a similar future save for some ``anomalies''. This motivates us to make use of all available historical data across users for prediction.

SubseqLS predicts one-day ahead by first setting the target account's balance sequence from the immediate past as the length$-L$ query vector $Y$. It then selects $M$ balance sequences $\left\{f^{(m)}\right\}_{m=1}^M$ that are similar to $Y$ in its first $L_1\leq L$ values from historical balances of all users. Finally, it determines the best weights for the $M$ sequences to match $Y$. We combine the $M$ sequences linearly for simplicity of the model, but more flexible combinations are also possible in principle. Essentially, for account $A$ of user $U$, we consider
$$ Y_\ell = \alpha_0 + \sum_{m=1}^M \alpha_m \left[h_{A,U}\left( f^{(m)} \right)\right]_\ell$$
for some transformation $h_{A,U}$ and estimate the coefficients $\alpha$ for a good prediction performance. Weights determined in traditional KNN regression methods tend to overfit to the query and are not robust to ``anomalies'', so we regularize the estimation to avoid this.

Let $t$ be the current date, and $S=31$ be the number of days to predict. We set $L=31$ so that $Y$ is sufficiently long to capture most recurring transactions. To recap notations, $Y=[Y_{t-L+1},\dots,Y_t]^T$ is the query for some account $A$ of some user $U$, and $Y_{t+s}$ is the $s$-day ahead balance to be predicted. Each selected sequence $f^{(m)}= \left[f^{(m)}_{t-L+1},\dots,f^{(m)}_{t+S}\right]^T$ is $L+S$ in length.
We use DTW distance with window = 2 to measure sequence similarity to allow slight misalignments, and iteratively find each $f^{(m)}$ using the fast search in \cite{rakthanmanon12}. We then locally expand or contract the sequences to adjust for temporal variations through function $h_{A,U}$, which aligns each $f^{(m)}$ to a template of payday events, since paydays mark the start of cyclic spending patterns. Payday estimation is described in Section \ref{sec: recur}.
Denoting the aligned sequence as $\left[\tilde{f}^{(m)}_{t-L+1},\dots,\tilde{f}^{(m)}_{t+S}\right]^T = \tilde{f}^{(m)} = h_{A,U}\left(f^{(m)}\right)$, the first subsequence with length $L$ is used to match $Y$, and the second with length $S$ is used for prediction. Note that $\tilde{f}^{(m)}_\ell$ may not be observed at time $\ell$ but is just matched to $Y_\ell$. 

We outline SubseqLS algorithm for one-day ahead prediction below. All sequences mentioned are standardized.
\begin{enumerate}
\item Set query vector $Y$ consisting of user U, account A's daily balance from $t-L+1$ to $t$.
\item Find $M$ sequences $\left\{f^{(m)}\right\}_{m=1}^M$ of length $L+S$ most similar to $Y$ in DTW distance in the first $L_1$ values.
\item Create a template of indicators marking user U's paycheck deposits into account A between $t-L+1$ and $t+S$, with magnitude being the paycheck values. Set $h_{A,U}$ to be the function that aligns any sequence to this template by DTW.
\item Align each $f^{(m)}$ such that $\tilde{f}^{(m)} = h_{A,U}\left(f^{(m)}\right)$.
\item Estimate $\beta_0$ and nonnegative $\beta$ coefficients which minimize the following objective
$$ \sum_{\ell=t-L+1}^{t}  w_\ell \left(Y_\ell - \beta_0 - X_\ell \beta\right)^2  + \lambda \|D\beta\|^2, $$ where $X_\ell = \left[\tilde{f}^{(1)}_\ell, \dots, \tilde{f}^{(M)}_\ell\right] $,   $D_{i,j} = \left|\tilde{f}^{(i)}_{t+1} - \tilde{f}^{(j)}_{t+1}\right|$ and obtain $\hat{\beta}_0$ and $\hat{\beta}$.

\item Estimate $\widehat{Y}_{t+1} = \hat{\beta}_0 + \left[\tilde{f}^{(1)}_{t+1}, \dots, \tilde{f}^{(M)}_{t+1}\right]\hat{\beta}$.

\end{enumerate}

\begin{figure}
\setlength{\belowcaptionskip}{-5pt}

    \begin{subfigure}[b]{0.48\linewidth}
        \centering
        \includegraphics[width=\linewidth]{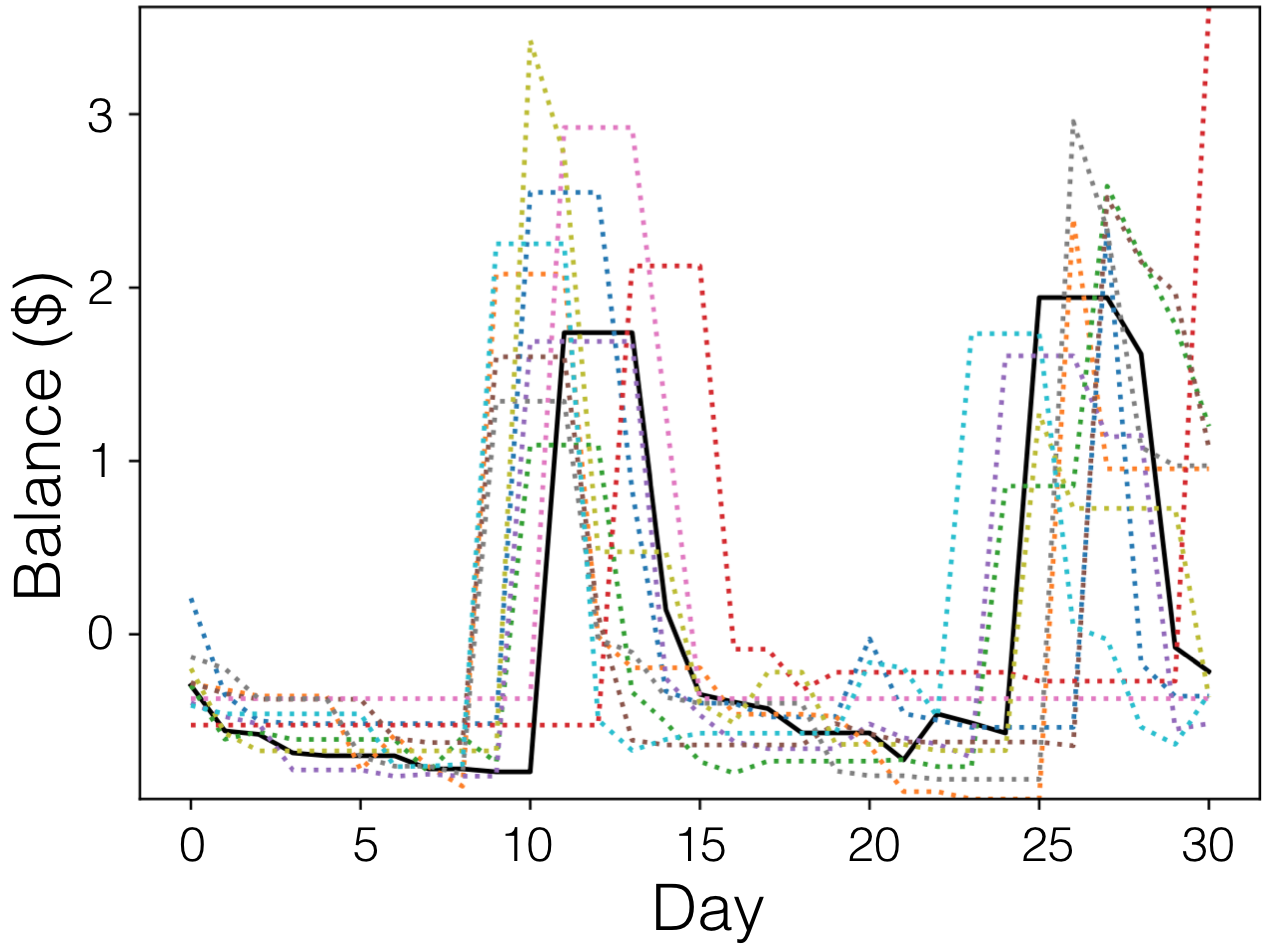}
        \caption{Original sequences}
    \end{subfigure}
    \hspace{\fill}
    \begin{subfigure}[b]{0.48\linewidth}
        \centering
        \includegraphics[width=\linewidth]{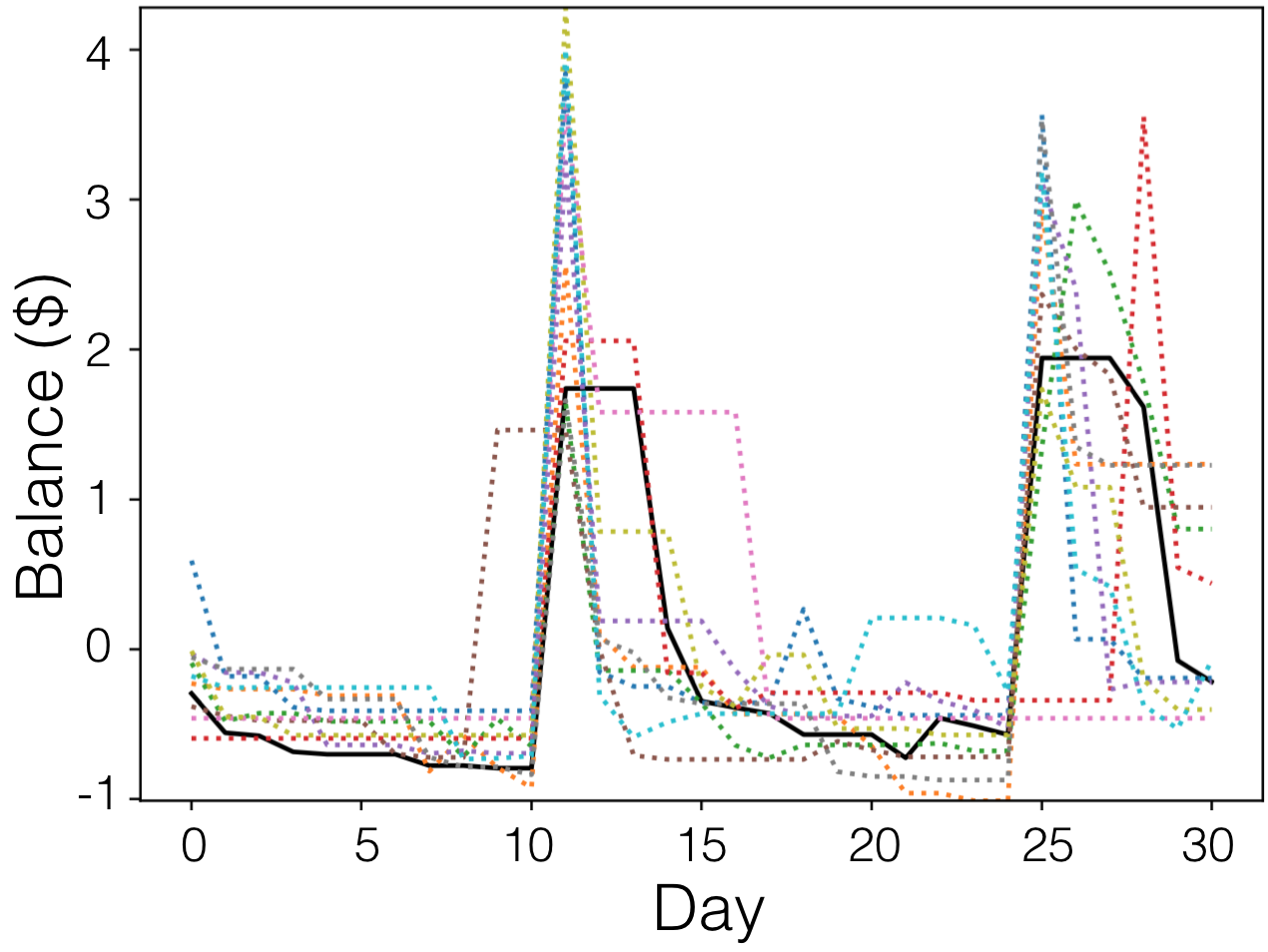}
        \caption{Aligned sequences}
    \end{subfigure}
    
    \caption{Effect of aligning to template of payday landmarks: black solid line is query $Y$, colored dotted lines are first $L$ values of matched sequences.}
    \label{fig: align}
\end{figure}

Aligning in Step 4 is an essential adjustment for temporal variations of matched sequences. Figure \ref{fig: align} shows how this preserves the exact cyclic pattern of $Y$.

We set $L_1=20$ in Step 2 so that similarity matching is done on a sufficiently long sequence with at least one semimonthly pay period. We let $L_1<L$ to eliminate sequences in $\left\{f^{(m)}\right\}_{m=1}^M$ which do not consistently match $Y$ in Step 5. Matrix $D$ also penalizes $\beta$ based on anomalous predictions of each $f^{(m)}$. Weight $w_\ell$ is 1 if $\ell\leq L_1$, 5 if $L_1<\ell< L$ and 10 if $\ell=L$ to emphasize more accurate estimation of the tail of $Y$ which is closer to the start of prediction.

\subsection{Extraction of recurring transactions and unexpected large expenses}
\label{sec: recur}

\begin{figure}
\setlength{\belowcaptionskip}{-10pt}
\includegraphics[width=\linewidth]{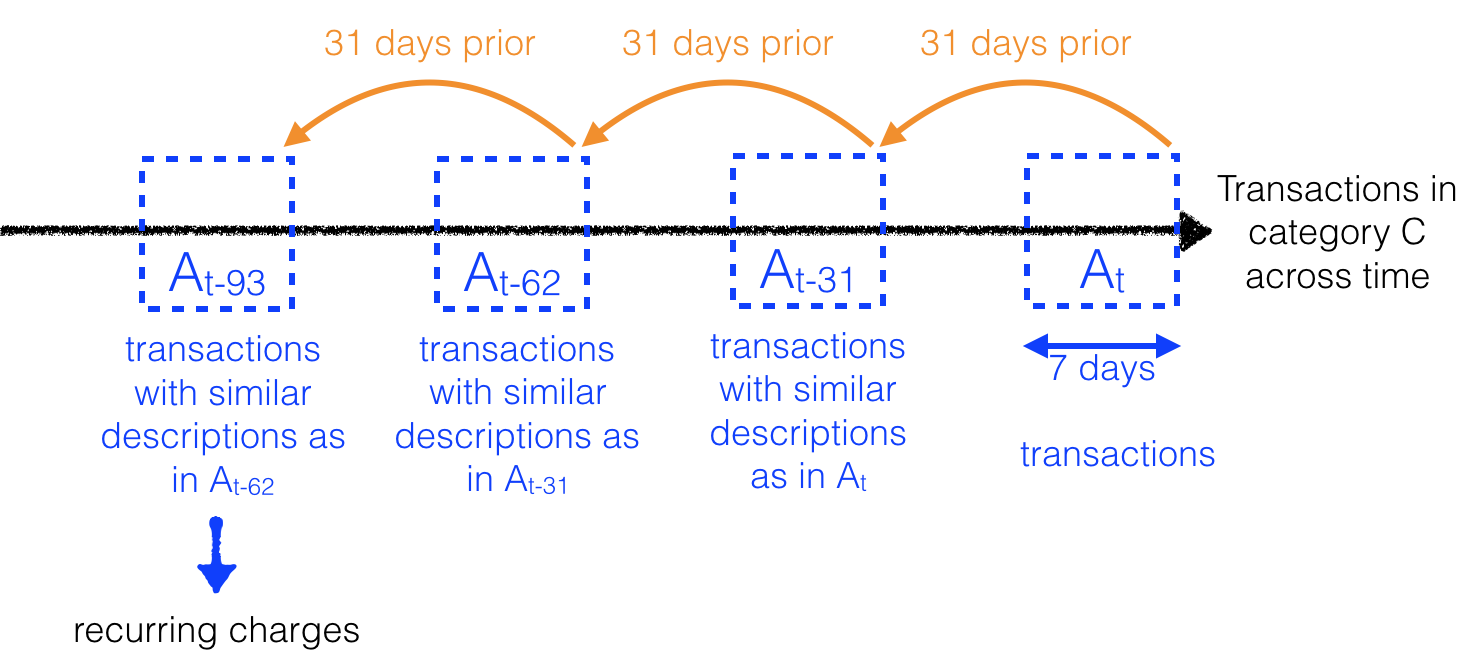}
\caption{Procedure to extract monthly charges.}
\label{fig: recur}
\end{figure}

We propose a procedure to automatically extract recurring transactions in each account, which include bills and periodic behavior such as salary, recurring money transfer, and grocery shopping. They are split into monthly, semimonthly, biweekly and weekly frequency. For spending category C, for each transaction and frequency, we look for past transactions with similar descriptions satisfying the frequency constraint. 
Figure \ref{fig: recur} shows the procedure for monthly charges. We start with all transactions within a 7-day window, denoted $A_{t}$. We backtrack by 31 days and retrieve all transactions in a window that have descriptions similar to those in $A_{t}$. We repeat till we obtain 4 windows of transactions to ensure that the remaining transactions identified in the last window indeed have the desired frequency.

For monthly and semimonthly charges, we use a 7-day window to accommodate differing month lengths, and use a 2-day window otherwise to accommodate small spending shifts due to holidays, etc. To compare descriptions, we use the Python \texttt{difflib} module \cite{difflib} to iteratively find the longest contiguous matching subsequences excluding junk elements, with a similarity threshold ratio of $0.75$ to accommodate insignificant differences such as dates and reference numbers.

To predict the next occurrence of a monthly transaction, we estimate the date as the last observed transaction date plus one month, and the amount as the historical average. We do similarly for semimonthly, biweekly and weekly transactions. 

We further use the recurring transactions to find unexpected or anomalous large expenses. On each user's transactions:
\begin{enumerate}
\setlength\itemsep{0em}
\item Remove all recurring transactions;
\item Retain unique transactions in remaining top 10\%.
\end{enumerate}

The results are pooled across all users, and the list of expenses displayed to a user can be personalized depending on their characteristics (e.g., car owner, a person with children).

\section{Empirical Evaluation}

\subsection{Data Description}
\label{sec: desc}

Since WageGoal is a relatively new app, the dataset collected is limited in terms of the number of users and the length of usage. We hence include an additional financial dataset from the PKDD'99 Discovery Challenge. We note that the PKDD'99 Dataset is not specific to low-wage workers.

\subsubsection{WageGoal Dataset}
\label{sec: wagegoaldata}

This dataset is collected from 19 individuals, with approximately one year's worth of financial transactions from June 21, 2016, to June 16, 2017, in checking, savings, and credit card accounts. There are a total of 52 accounts of which 16 are checking, 19 are savings, and 17 are other accounts including credit cards. Each line item in the data includes date, description, amount, category and final account balance as described in Section \ref{sec: overview}. The one year's worth of data is split into a training period of nine months and a testing period of three months. All users have semimonthly income.

\subsubsection{PKDD'99 Dataset}

This is a publicly available dataset of real anonymized bank transactions from January 1, 1993 to December 31, 1998 \cite{pkdd99}. We test here the scenario of long historical data and retain the 2263 accounts with at least four years of data. Accounts have sparse transactions with a maximum of 52.479 per year, so we consider weekly instead of daily balances. Training and testing periods are 4.5 and 1.5 years respectively. Each line item includes date and amount. No information on description and category is provided, and hence this dataset is not used to evaluate any other task besides prediction. 

\subsection{Prediction of account balances}

From the test set, 25 length-31 sequences are randomly chosen for prediction. We compute two different error measures:
\begin{itemize}
\setlength\itemsep{0em}
\item MAE (Mean Absolute Error), the average absolute difference between true and predicted account balances;
\item Average difference in dollar amounts between true and predicted balances when the former becomes negative.
\end{itemize}
The point in time at which balances become negative is of special interest because penalty fees will start being charged. We use only the first error measure on the PKDD'99 dataset, the second is not applicable because true account balances in the PKDD'99 dataset are unknown and we arbitrarily set all account balances to start at 0.
To calculate the error measures, we scale all accounts to have variance equal to 100 so that they contribute approximately equally.

We compare the performance of the individual methods HistAvg and SubseqLS, the hybrid HistAvg-SubseqLS, as well as Prophet, ARMA, NearestNeighbor and KNN averaging. 
Prophet is a state-of-the-art forecasting method from Facebook \cite{taylor17} that uses a regression model to fit a linear trend, and incorporates weekly seasonality and holiday effects by marking them through indicator variables. We used paydays in lieu of holiday effects.
ARMA is a well-established model for stationary stochastic processes \cite{shumway06}, and we implemented ARMA with parameters found through the \texttt{statsmodels} module using default arguments \cite{statsmodels}.
K-Nearest Neighbor is a popular nonparametric approach that is flexible and applied widely in diverse domains such as traffic flow and energy \cite{alvarez11,zhang13}. In NearestNeighbor, only the top matched sequence to the query is used by directly taking the value immediately after the match as the one-day-ahead prediction, and in KNN averaging, the top 10 matched sequences are used and the one-day-ahead prediction is the average of the values immediately after all 10 matches.

\subsubsection{Prediction with WageGoal Dataset}
\label{sec: pred_wagegoal}

\begin{figure}
\setlength{\belowcaptionskip}{-2pt}

    \begin{subfigure}[b]{0.48\linewidth}
        \centering
        \includegraphics[width=\linewidth]{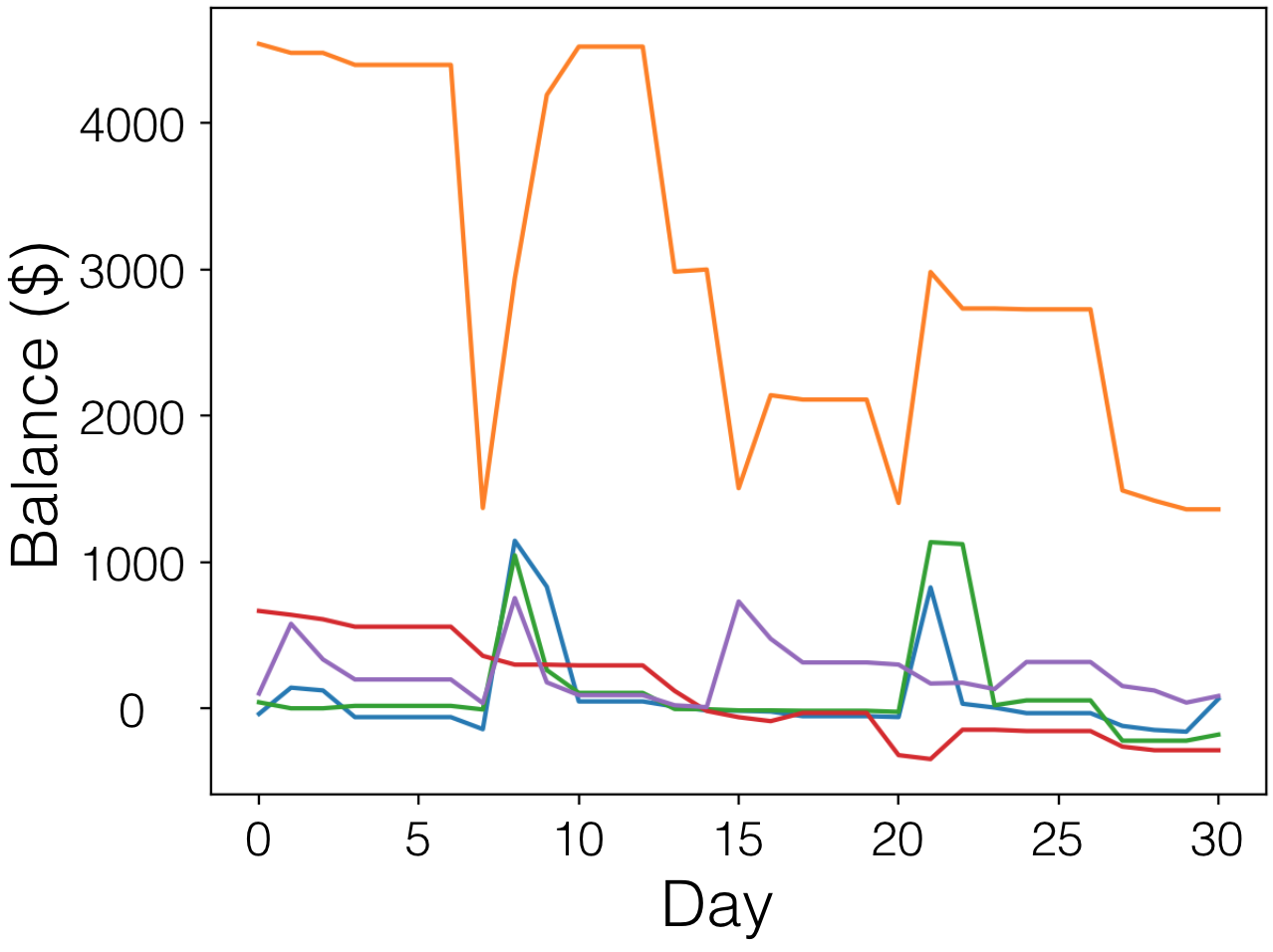}
        \caption{Example balance sequences}
    \end{subfigure}
    \hspace{\fill}
    \begin{subfigure}[b]{0.48\linewidth}
        \centering
        \includegraphics[width=\linewidth]{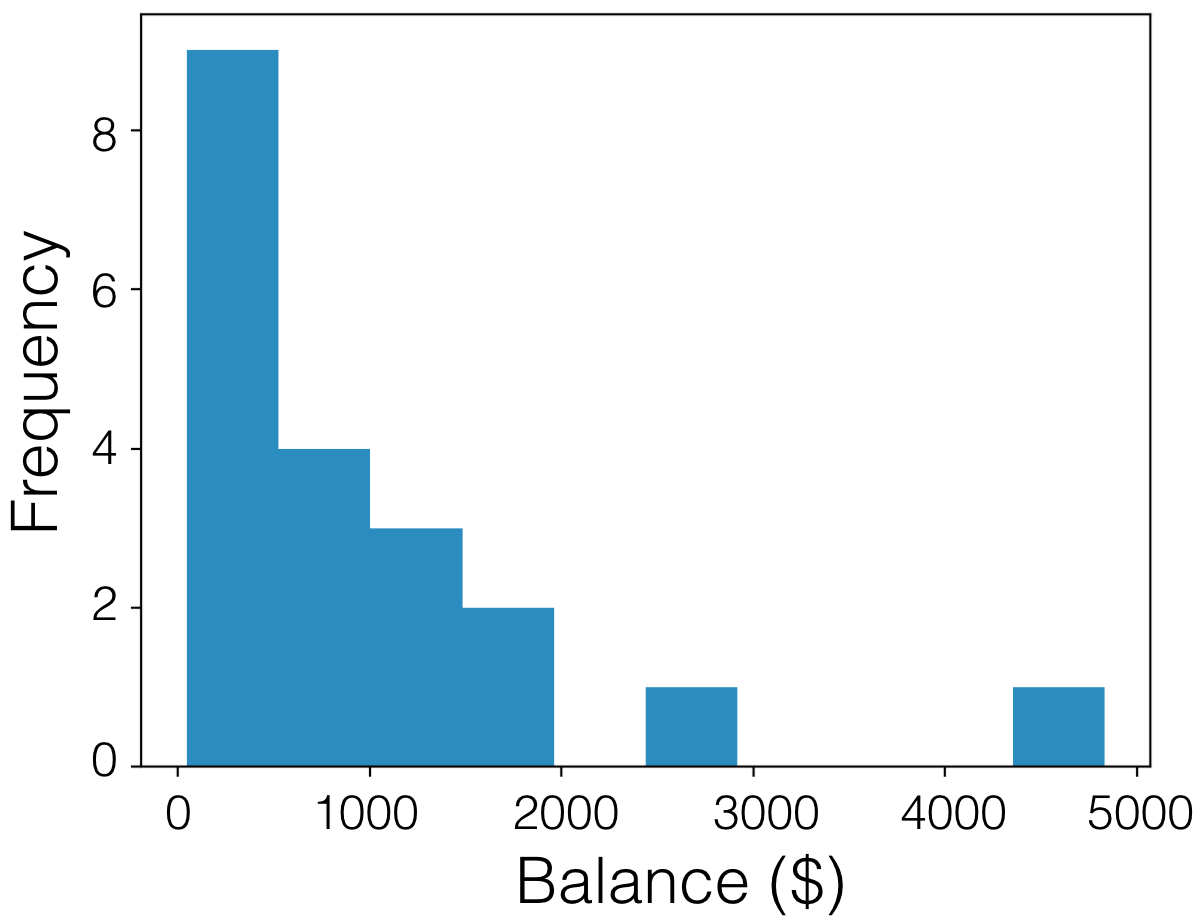}
        \caption{Standard deviations}
    \end{subfigure}
    \caption{WageGoal Dataset: Paycheck accounts}
    \label{fig: wagegoal_pay}

    \begin{subfigure}[b]{0.48\linewidth}
        \centering
        \includegraphics[width=\linewidth]{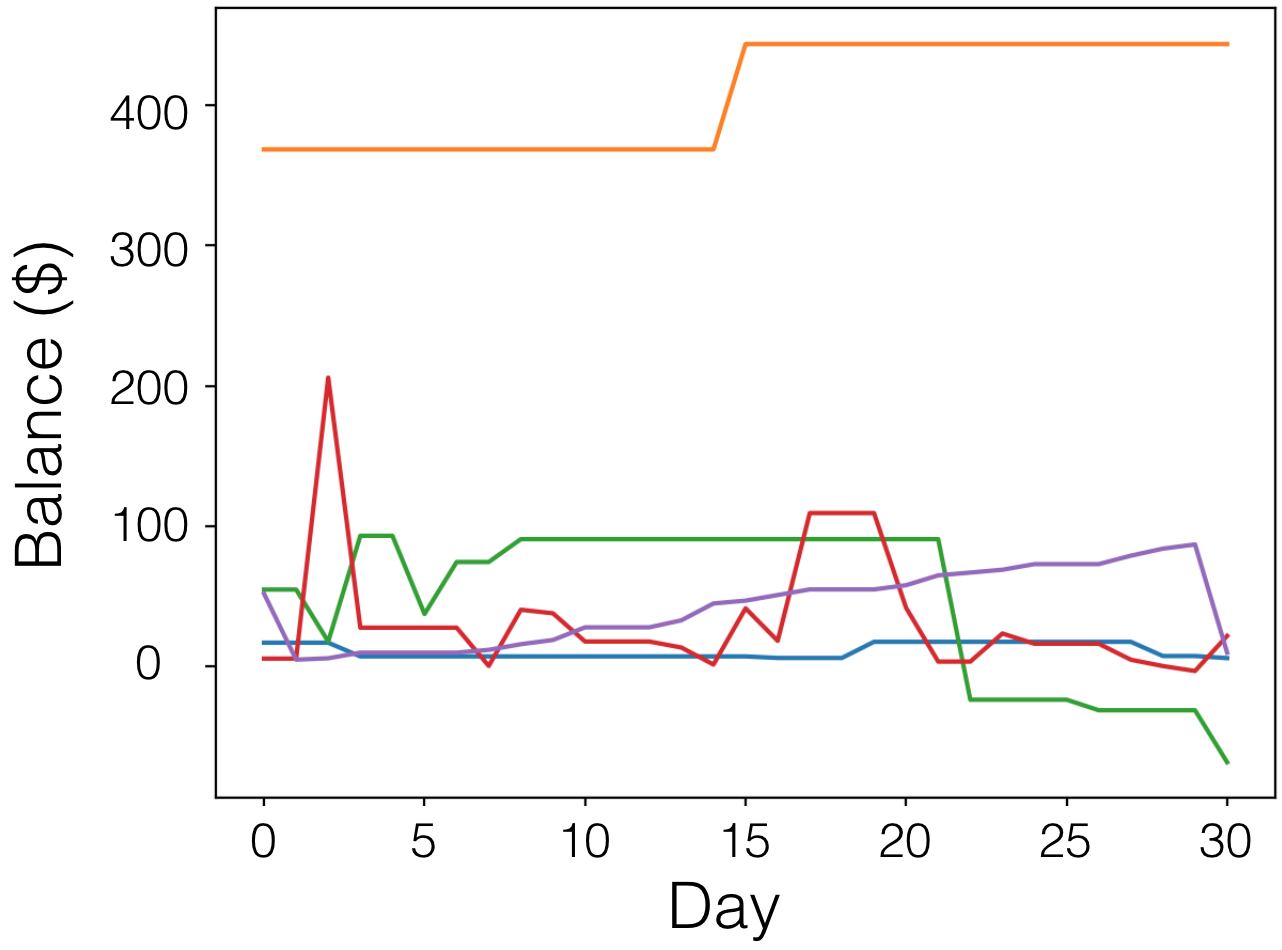}
        \caption{Example balance sequences}
    \end{subfigure}
    \hspace{\fill}
    \begin{subfigure}[b]{0.48\linewidth}
        \centering
        \includegraphics[width=\linewidth]{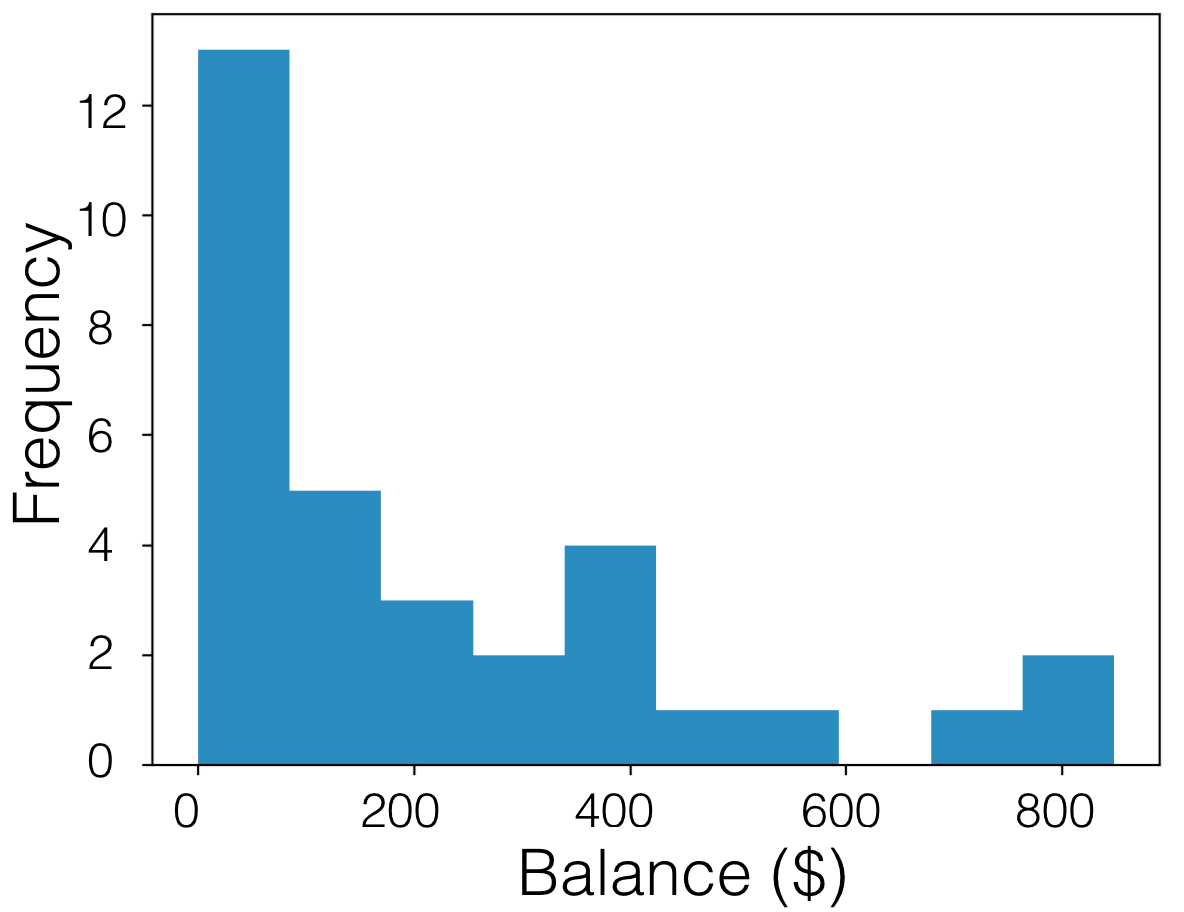}
        \caption{Standard deviations}
    \end{subfigure}
    \caption{WageGoal Dataset: Non-paycheck accounts}
    \label{fig: wagegoal_nopay}
    
\end{figure}

The WageGoal Dataset is split into two types of accounts, those with paycheck income transactions (20 accounts), and those without (32 accounts). The former demonstrates more pronounced cyclic pattern as in Figure \ref{fig: wagegoal_pay} and \ref{fig: wagegoal_nopay}.

The training set is used to optimize the number of matches $M$ and penalty parameter $\lambda$ in SubseqLS by cross-validation and also to determine the switching parameter $\tau$ for HistAvg-SubseqLS. Search range for $M$ is in multiples of 5 between 5 and 25, $\lambda$ values are between 0 and 10, and $\tau$ values are integers between 0 and $S=31$. For paycheck accounts, $M_{pay}=10$ and $\tau_{pay}=3$. For non-paycheck accounts, $M_{nopay}=5$ and $\tau_{nopay}=31$, meaning HistAvg-SubseqLS reduces to HistAvg. Parameter $\lambda$ is determined individually for each account and hence not reported here.

\begin{figure}[h]

    \begin{subfigure}[b]{\linewidth}
        \centering
        \includegraphics[width=0.7\linewidth]{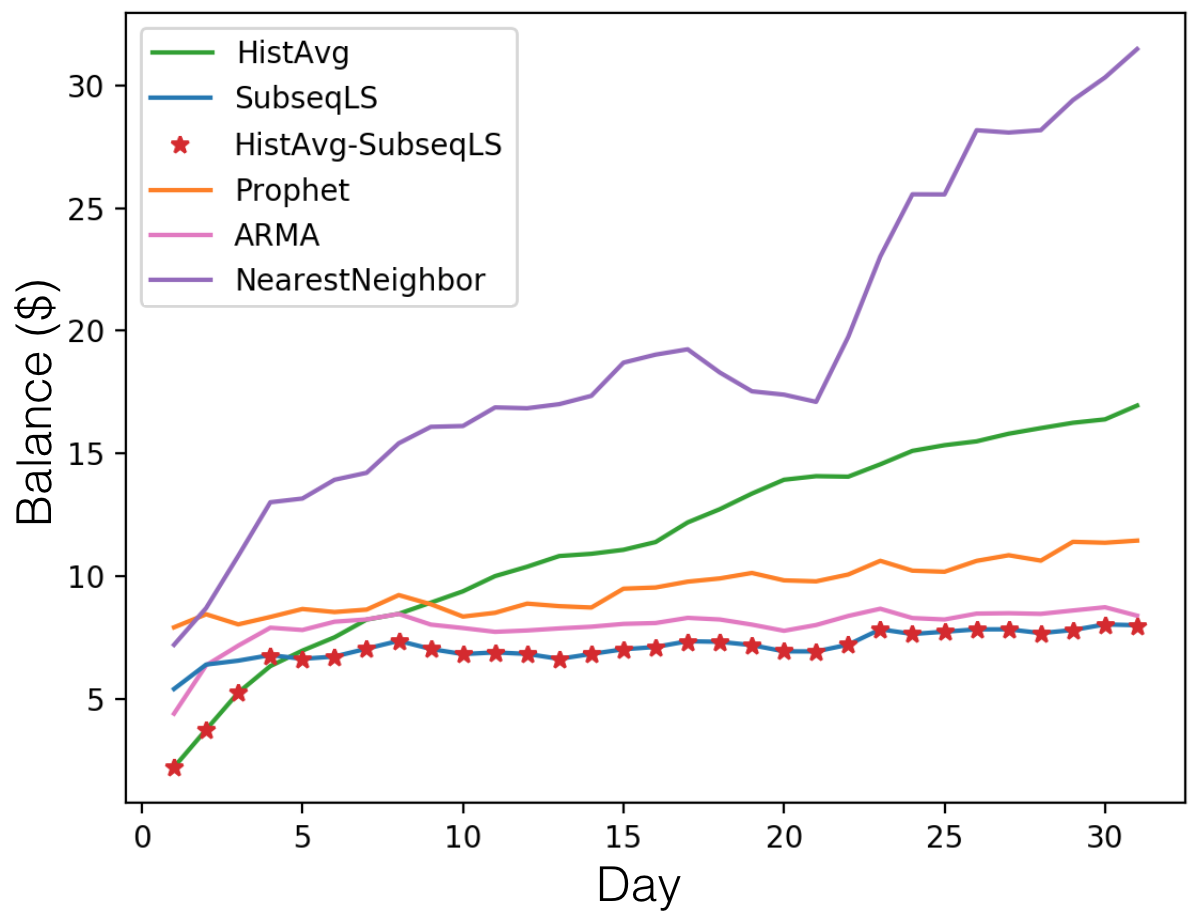}
        \caption{Paycheck accounts}
        \label{fig: pred_pay}
    \end{subfigure}
    
    \begin{subfigure}[b]{\linewidth}
        \centering
        \includegraphics[width=0.7\linewidth]{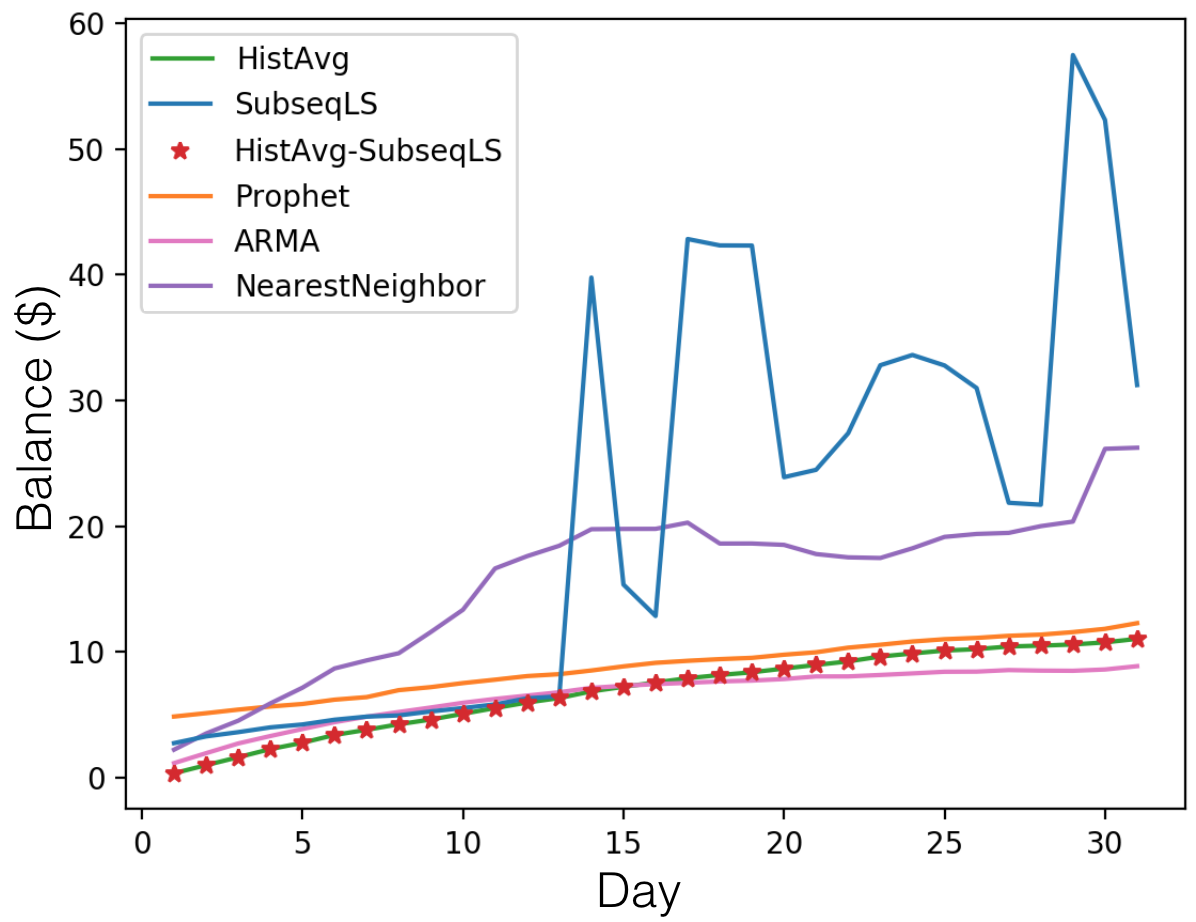}
        \caption{Non-paycheck accounts}
        \label{fig: pred_nopay}
    \end{subfigure}
    
    \caption{WageGoal Dataset: Average account balance prediction error over time across 25 test samples. KNN is excluded due to large magnitude of error.}
    \label{fig: pred}
\end{figure}

Table \ref{table: pred} shows the test results. HistAvg-SubseqLS almost always performs the best, and is otherwise a close second. 
Figure \ref{fig: pred} plots the average absolute difference between the actual and predicted account balance across time. 

Paycheck account balances tend to have pronounced semimonthly patterns starting with a sharp increase at payday followed by a decrease to pre-payday levels. These cyclic patterns sometimes perpetuate through historical data and are shared across users, which make sequence-matching suitable. SubseqLS benefited from regularization in this small dataset because not all top matches were close matches. KNN, in contrast, had average prediction error at least 50 times higher than all other methods.
In Figure \ref{fig: pred_pay}, SubseqLS maintained almost consistent error across time, while other methods had higher errors predicting further ahead. Due to the regression formulation, SubseqLS focuses on modeling overall trend instead of next-day prediction. Weights $w$ provides some balance, but are difficult to tune. In Figure \ref{fig: pred_pay}, HistAvg had the best short-term predictions as its next-day prediction is designed to be close to the current observation. Switching parameter $\tau_{pay}$ let HistAvg-SubseqLS use HistAvg for 3 days before switching to SubseqLS, thereby attaining the lowest errors in both short- and long-term.

Non-paycheck accounts are mostly used for savings and occasional purchases. Common transactions include spare change saved through bank programs. The lack of prominent structures in balance sequences resulted in poor matches found for NearestNeighbor, KNN and SubseqLS. As in Figure \ref{fig: pred_nopay}, HistAvg performed the best by making conservative predictions. The switching parameter $\tau_{nopay}$ correctly picked to use HistAvg throughout the prediction period, so that HistAvg-SubseqLS shared the same good performance.

\begin{table}

\begin{subtable}{.5\textwidth}
\centering
\begin{tabular}{l*{2}{c}}
Methods              & MAE & Error in amount \\
\hline
HistAvg & 11.417 & 7.460 \\
SubseqLS    & \underline{7.005} & \underline{5.774} \\ 
HistAvg-SubseqLS  & \textbf{6.790} & \textbf{5.099} \\
Prophet & 9.534 & 7.508 \\
ARMA & 7.941 & 6.983 \\
NearestNeighbor & 19.126 & 11.851 \\
KNN & 1105.719 & 88.193 \\
\end{tabular}
\caption{Paycheck accounts}
\end{subtable}

\begin{subtable}{.5\textwidth}
\centering
\begin{tabular}{l*{2}{c}}
Methods              & MAE & Error in amount \\
\hline
HistAvg & \underline{6.876} & \underline{6.863} \\
SubseqLS    & 18.474 & 7.226 \\
HistAvg-SubseqLS  & \underline{6.876} & \underline{6.863} \\
Prophet & 8.794 & 11.944 \\
ARMA & \textbf{6.565} & 6.981 \\
NearestNeighbor & 15.439 & \textbf{6.832} \\
KNN & 7212.774 & 370.220 \\
\end{tabular}
\caption{Non-paycheck accounts}
\end{subtable}

\caption{WageGoal Dataset: Account balance prediction errors averaged across 25 one-month test samples.}
\label{table: pred}
\vspace{-10pt}
\end{table}

\subsubsection{Prediction with PKDD'99 Dataset}

\begin{figure}
\setlength{\belowcaptionskip}{-10pt}

    \begin{subfigure}[b]{0.48\linewidth}
        \centering
        \includegraphics[width=\linewidth]{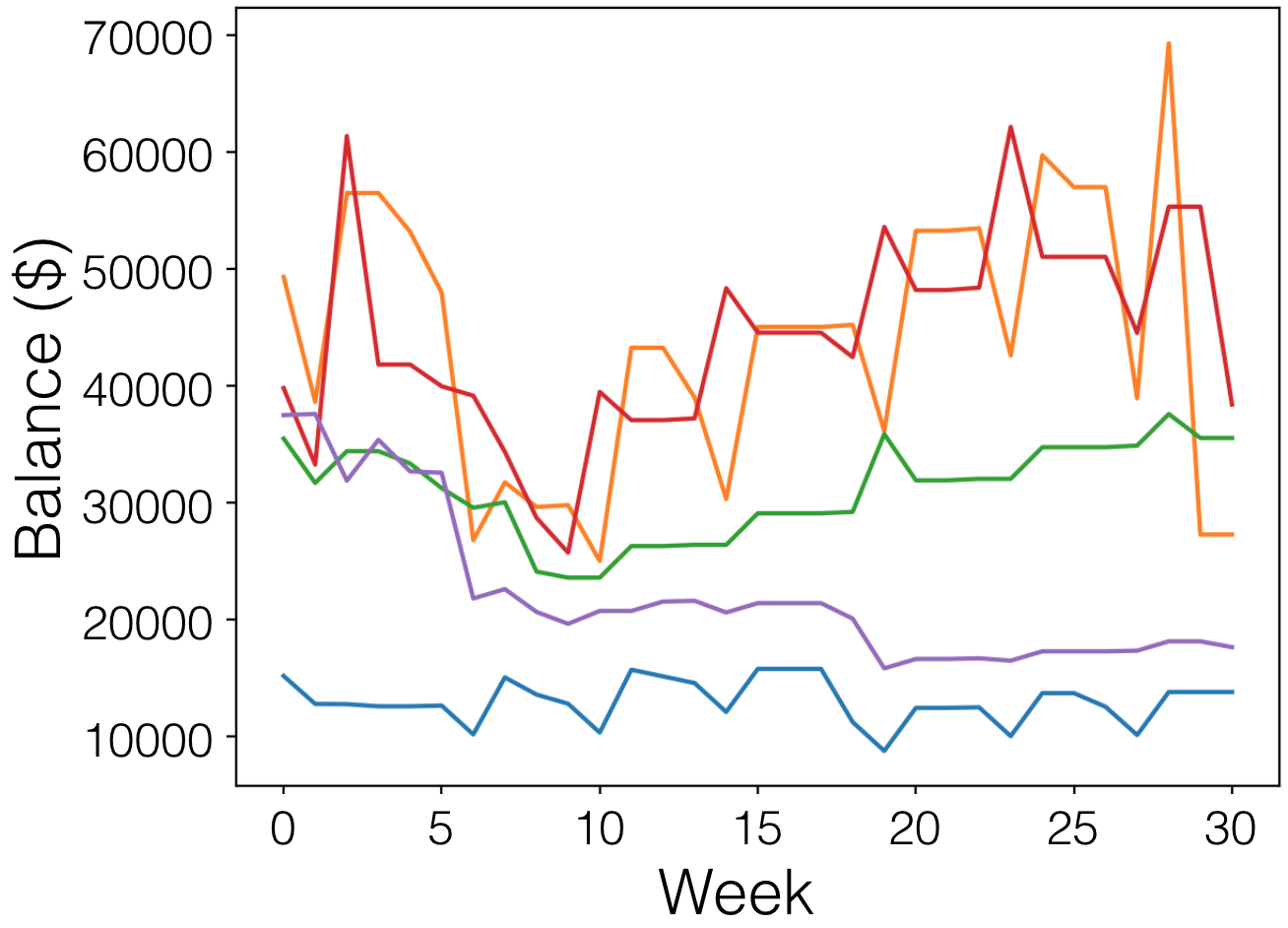}
        \caption{Example balance sequences}
    \end{subfigure}
    \hspace{\fill}
    \begin{subfigure}[b]{0.48\linewidth}
        \centering
        \includegraphics[width=\linewidth]{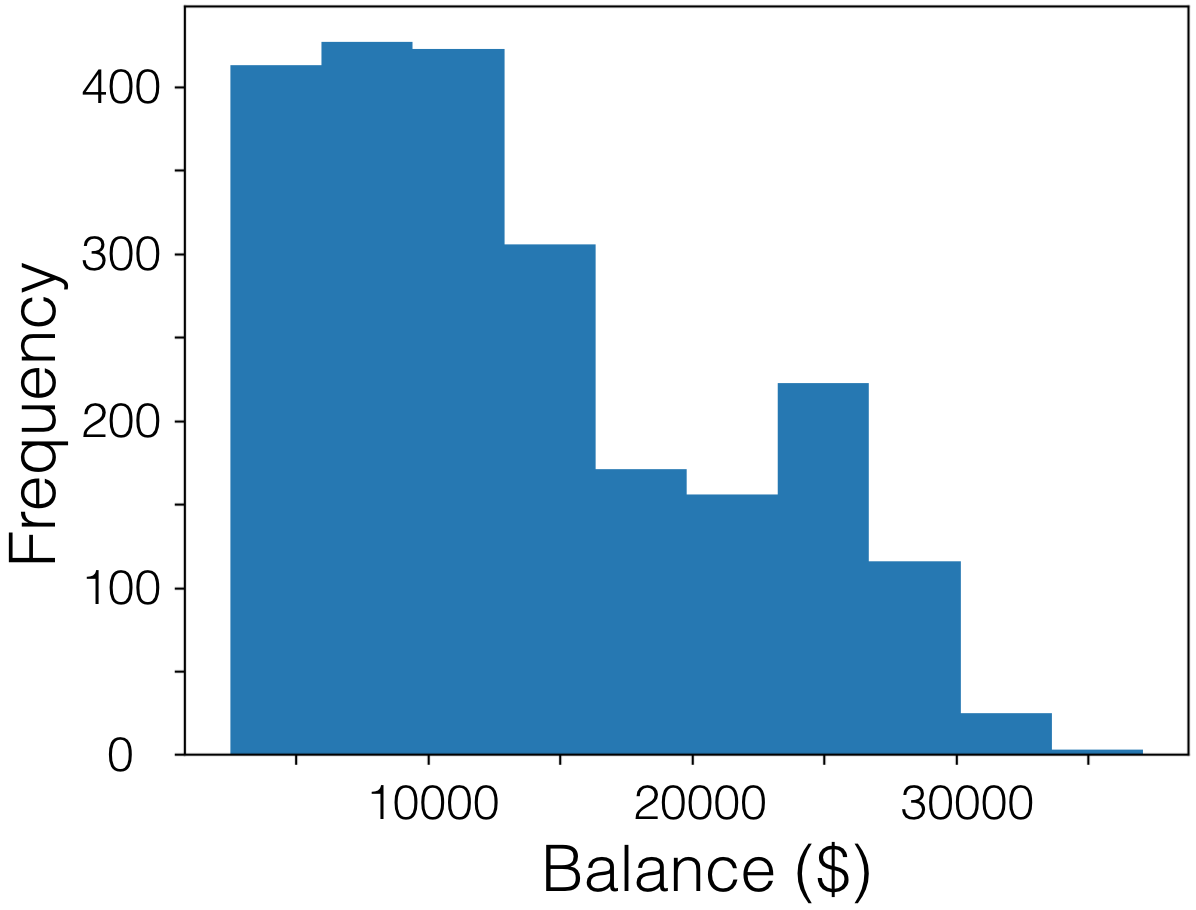}
        \caption{Standard deviations}
    \end{subfigure}
    \caption{PKDD'99 Dataset: Properties of accounts}
    \label{fig: pkdd}
    
\end{figure}

Balances exhibit cyclic patterns as in Figure \ref{fig: pkdd}, just like WageGoal paycheck account balances. Since transaction descriptions and categories are not available, we cannot extract any recurring transaction for HistAvg, SubseqLS, and Prophet. We also do not report HistAvg-SubseqLS, since HistAvg, being reliant on good estimates of recurring transactions, is heavily handicapped in this setup and does not benefit the hybrid approach.

\begin{figure}
\setlength{\belowcaptionskip}{-10pt}

    \centering
    \includegraphics[width=0.7\linewidth]{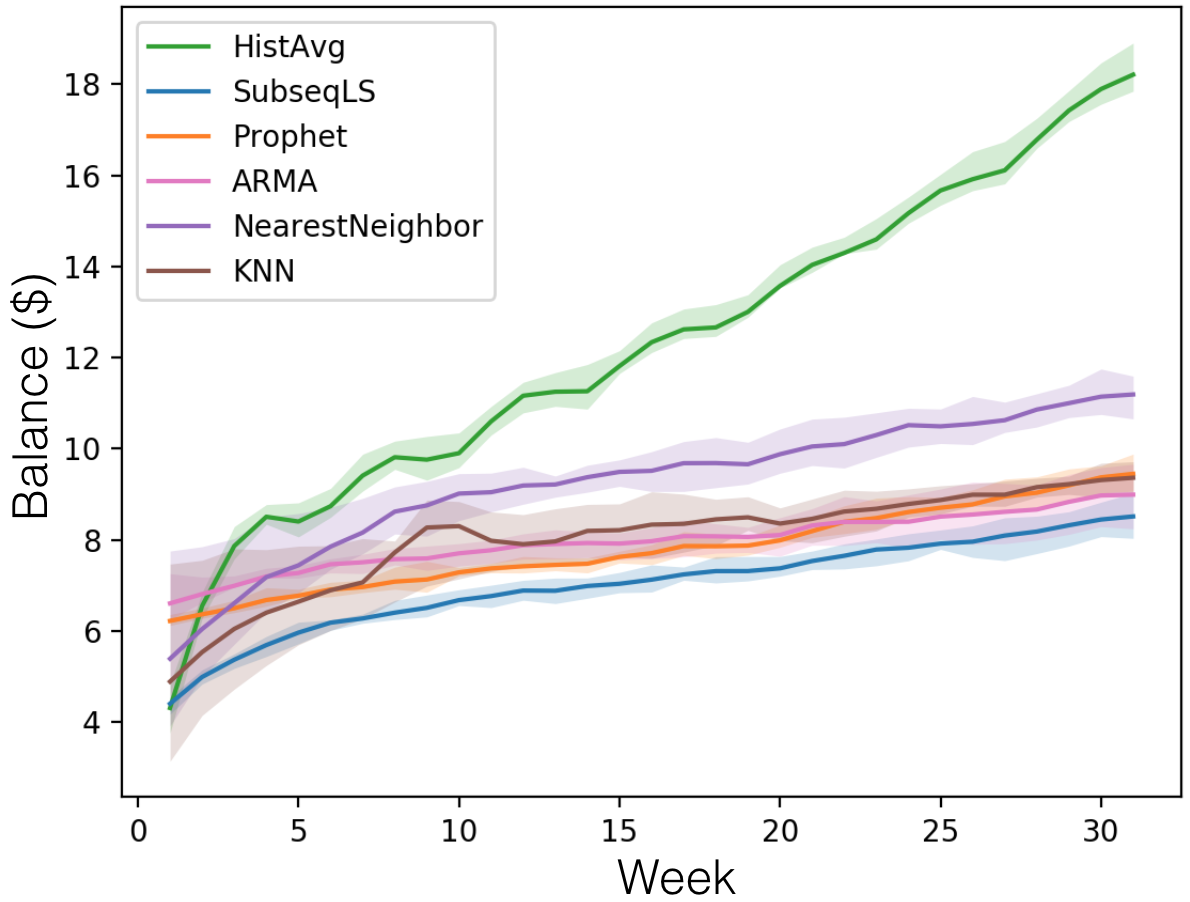}
    
    \caption{PKDD'99 Dataset: Average account balance prediction error over time across 25 test samples.}
    \label{fig: pred_pkdd}
\end{figure}

In each iteration of the experiment, we randomly select 52 accounts to predict. We present test results across eight iterations in Table \ref{table: pkdd_pred}. Figure \ref{fig: pred_pkdd} plots the average absolute difference between the actual and predicted account balance across time. Solid lines are the mean across all iterations, and the shaded regions are the $25^{th}$ to $75^{th}$ percentile.

Results mirror those in Section \ref{sec: pred_wagegoal} that SubseqLS performs the best in long-term predictions, showing it is consistent across different scenarios and dataset sizes. Furthermore, in this larger dataset, SubseqLS achieved low errors in short-term predictions as well. Larger data benefit sequence-matching methods since more close matches are available. For instance, KNN's performance for the PKDD'99 Dataset is also much higher than that for the WageGoal Dataset.

\begin{table}

\centering
\begin{tabular}{l*{2}{c}}
Methods              & MAE Mean & MAE Standard deviation \\
\hline
HistAvg & 12.245 & 0.578 \\
SubseqLS & \textbf{7.017} & \textbf{0.277} \\ 
Prophet & \underline{7.794} & \underline{0.311} \\
ARMA & 7.967 & 0.455 \\
NearestNeighbor & 9.243 & 0.674\\
KNN & 8.012 & 0.751 \\
\end{tabular}

\caption{PKDD'99 Dataset: Account balance prediction errors averaged across 25 length-31 test samples.}
\label{table: pkdd_pred}
\vspace{-10pt}
\end{table}

\subsection{Extraction of recurring transactions and unexpected large expenses}
\label{sec: eval recur}

From the test set of the WageGoal Dataset, 25 dates are randomly picked. For each date, we extract recurring transactions prior to the date, and predict the dates for their next occurrence. A transaction is correctly extracted if the prediction is within 5 days of the true date. We evaluate the quality of the extraction procedure by:
\begin{itemize}
\setlength\itemsep{0em}
\item Average number of true recurring transactions extracted per user;
\item Precision, i.e., proportion of recurring transactions extracted that is true;
\item Average error in days for the predicted dates at which recurring transactions next occur.
\end{itemize}

We compare the performance of our proposed procedure with an extraction method utilizing transaction descriptions and category labels. The competing method flags a transaction as recurring if the description contains the word `recurring' or if the category label contains the following keywords: `bill pay', `payroll', `service - insurance' and `service - subscription'. These rules are manually formulated based on close inspection of the dataset.

As in Table \ref{table: recur}, the proposed method identified a larger number of true recurring transactions and with higher precision. Despite extracting more than double the number of true recurring transactions, the proposed method was only half a day worse on average in predicting the next transaction date. We combine the recurring transactions found by both methods above and use them to obtain a list of unexpected large expenses. Some examples of their approximate costs are shown in Table \ref{table: large}. We provided only a single value for each cost, but given observations of the expense from more users, a range or empirical distribution will be appropriate. 

\begin{table}

\centering
\begin{tabular}{l*{3}{c}}
Method & \# extracted & Precision & \makecell{Error in \\days} \\
\hline
Proposed & \textbf{4.633} & \textbf{0.647} & 1.465 \\
Using labels & 2.161 & 0.311 & \textbf{1.043} \\
\end{tabular}

\caption{WageGoal Dataset: Performance in extraction of recurring transactions across 25 test samples.}
\label{table: recur}
\vspace{-10pt}
\end{table}


\begin{table}

\centering
\begin{tabular}{l*{1}{c}}
Description & Cost (\$) \\
\hline
House cleaning service & 350 \\
Car repair & 500 \\
Student loan & 5000 \\
Roofing & 8000 \\
\end{tabular}

\caption{WageGoal Dataset: Examples of unexpected large expenses.}
\label{table: large}
\vspace{-10pt}
\end{table}

\section{Significance and Impact}

Our system will upgrade and replace existing models in WageGoal, and therefore have a direct and near-immediate impact on the low-income individuals connected to the product. 
The enhancements will help users manage their volatile cash flow, capitalize on opportunities for debt reduction and savings, obtain greater overall financial health and stay out of poverty. 
Strategic partnerships will help Neighborhood Trust further penetrate relevant markets in the coming years, eventually reaching many tens of thousands of clients nation-wide through its technology platforms.

Robust tracking systems are in place to measure the outcomes. Results will be shared as
appropriate via Neighborhood Trust's network of financial empowerment providers and other
interested stakeholders.

\section{Conclusion}

We proposed a system of data mining techniques to predict and analyze spending behaviors. Future work includes improving the predictive models by incorporating additional individual- and group-level information, providing early and enhanced visibility for users into their financial health, and automatically generating personalized recommendations for improving financial stability. Depending on deployment feedback, other improvements will be considered as necessary.

\section{Acknowledgments}

The authors thank Mary Coker, Camilla Nestor, and Steve Silverstein from Neighborhood Trust, Karan Bhatia from Philosphie, and Sa\v{s}ka Mojsilovi\'c from IBM Research for their help and support.

\bibliographystyle{abbrv}
\bibliography{cfa}

\end{document}